\documentstyle[emulateapj,pstricks,psfig]{article}

\newcommand{\beqa}{\begin{eqnarray}}
\newcommand{\eeqa}{\end{eqnarray}}

\def\hMsun{\ h^{-1}M_{\odot}}

\def\gsim {\lower .1ex\hbox{\rlap{\raise .6ex\hbox{\hskip .3ex
        {\ifmmode{\scriptscriptstyle >}\else
                {$\scriptscriptstyle >$}\fi}}}
        \kern -.4ex{\ifmmode{\scriptscriptstyle \sim}\else
                {$\scriptscriptstyle\sim$}\fi}}}
\def\lsim {\lower .1ex\hbox{\rlap{\raise .6ex\hbox{\hskip .3ex
        {\ifmmode{\scriptscriptstyle <}\else
                {$\scriptscriptstyle <$}\fi}}}
        \kern -.4ex{\ifmmode{\scriptscriptstyle \sim}\else
                {$\scriptscriptstyle\sim$}\fi}}}
\def\beq{\begin{equation}}
\def\eeq{\end{equation}}
\def\ks{\rm ~km~s^{-1}}

\def\Rvir{R_{\rm vir}}
\def\Mvir{M_{\rm vir}}

\def\zre{z_{\rm re}}

\begin{document}
\slugcomment{{\em Astrophysical Journal, submitted}}
\lefthead{HIERARCHICAL GALAXY FORMATION AND SUBSTRUCTURE IN THE STELLAR HALO}
\righthead{BULLOCK, KRAVTSOV, \& WEINBERG}

\title{HIERARCHICAL GALAXY FORMATION AND SUBSTRUCTURE IN THE GALAXY'S STELLAR 
HALO}\vspace{3mm}

\author{James S. Bullock\altaffilmark{1}, Andrey V. Kravtsov\altaffilmark{2} 
and David H. Weinberg}
\affil{Department of Astronomy, The Ohio State University,
    140 W. 18th Ave, Columbus, OH 43210-1173}

\altaffiltext{1}{james,andrey,dhw@astronomy.ohio-state.edu}
\altaffiltext{2}{Hubble Fellow}

\begin{abstract}
  
We develop an  explicit model for the formation   of the stellar  halo
from  tidally disrupted, accreted  dwarf satellites  in the cold  dark
matter  (CDM) framework,  focusing on  predictions   testable with the
Sloan Digital Sky Survey (SDSS) and other wide-field surveys.  Subhalo
accretion and orbital  evolution are calculated using  a semi-analytic
approach  based on the  extended Press-Schechter formalism.  Motivated
by our  previous work,  we assume that  low-mass  subhalos ($v_c  < 30
\ks$) can form significant populations of  stars only if they accreted
a substantial fraction of their mass before the epoch of reionization.
With  this assumption,   the model  reproduces the observed   velocity
function  of  galactic  satellites in  the  Local Group,  solving  the
``dwarf satellite problem'' without modifying  the basic tenets of the
popular  $\Lambda+$CDM  cosmological  scenario.  The tidally disrupted
satellites in this model yield a stellar distribution whose total mass
and radial density profile are consistent  with those observed for the
Milky  Way stellar halo.  Most significantly,   the model predicts the
presence  of many  large-scale, coherent  substructures   in the outer
halo.  These  substructures   are  remnants   of  individual,  tidally
disrupted dwarf  satellite galaxies.   Substructure is more pronounced
at large galactocentric radii because of the smaller number density of
tidal streams  and  the longer orbital  times.  This  model provides a
natural explanation for the  coherent structures in the outer  stellar
halo found in  the SDSS commissioning  data, and it predicts that many
more such structures should be found as the survey  covers more of the
sky.   The detection  (or non-detection) and  characterization of such
structures  could eventually    test variants  of  the  CDM  scenario,
especially  those  that aim  to solve the  dwarf satellite  problem by
enhancing satellite disruption.

\end{abstract}

\section{Introduction}

The   origin   of the  Milky Way's   stellar  halo is  a long-standing
astronomical problem.   The poles  of  the debate are  defined by  the
monolithic collapse model of Eggen, Lynden-Bell, \& Sandage (1962) and
the   chaotic accretion  model of   Searle (1977)  and Searle \&  Zinn
(1978).  The  Searle \&  Zinn  picture has gained currency  in  recent
years  in part because of growing  recognition that the halo and bulge
are distinct components that  may have different  formation mechanisms
(see, e.g.,    the reviews by  Wyse  1999ab)  and in   part because of
``smoking gun'' evidence   that    includes the   tidally    distorted
Sagittarius dwarf galaxy (Ibata et  al.\ 2000a and references therein)
and the  presence  of extra-tidal  stars  around many dwarf spheroidal
satellites (Gould  et al.  1992;  Irwin \& Hatzidimitriou  1995; Kuhn,
Smith \&  Hawley 1996).   The  Searle \& Zinn  scenario  also bears  a
strong  anecdotal resemblance  to   the hierarchical galaxy  formation
scenario  characteristic  of   inflationary   cold  dark matter  (CDM)
cosmological  models.  In this  paper, we  make the connection between
CDM cosmology  and  hierarchical  stellar  halo formation   much  more
explicit, by presenting  a   simple but  quantitative model  for  halo
formation   in the CDM   framework  and obtaining  predictions for the
degree of residual substructure in the outer halo.

Previous  studies  of  stellar   halo  formation  in the  hierarchical
framework have focused on the fossil evidence for satellite disruption
preserved   in phase space    substructure of  the  halo stars  (e.g.,
Johnston, Spergel \&  Hernquist  1995; Helmi  \& White  1999).   These
studies were aimed primarily at exploiting surveys  of the halo in the
solar neighborhood  (e.g., Arnold \&  Gilmore 1992; Preston,  Beers \&
Shectman 1994;   Majewski, Munn \&  Hawley  1994,  1996).  Wide-angle,
deep, multi-color surveys, such as the Sloan Digital Sky Survey (SDSS;
York et al. 2000), open  up new avenues for  studying the structure of
the stellar  halo.  RR Lyrae stars, detected  by their variability and
color, can provide a three-dimensional map of the distribution of halo
stars.  The   more general population of A-colored   halo stars can be
used   for the same   purpose; relative   to RR  Lyrae  they have  the
advantages   of greater   numbers     and detectability in   a  single
observation epoch but the disadvantage  of being less precise standard
candles.  Studies  of  RR  Lyrae stars  and  A-colored  stars in  SDSS
commissioning  data have already  revealed  two large substructures in
the outer  halo (Ivezi\'{c}  et al.  2000; Yanny  et  al.  2000).  The
photometric depth of the SDSS and the intrinsic brightness of RR Lyrae
and A stars allows a probe  of halo structure  out to large distances,
$\sim 75$~kpc, and the  restricted absolute-magnitude range of RR Lyrae
and A stars prevents 3-dimensional substructure  from being washed out
by projection.  Carbon star  surveys (e.g., Ibata  et al.  2000a)  and
surveys of giant stars (Majewski et al. 2000) offer similar prospects.

The model presented in this paper is an extension of our
previous work aimed at explaining the observed abundance of dwarf
satellite galaxies in the Local Group within the CDM framework (Bullock,
Kravtsov \& Weinberg 2000; hereafter BKW).  In BKW we showed that the
observed shape and amplitude of the velocity function of dwarf
satellite galaxies in the Local Group can be explained if 
gas accretion and star formation are suppressed in low-mass dark matter
clumps after intergalactic gas
is reheated during the epoch of reionization.  In this picture, the
observed dwarf satellites around our Galaxy are those that assembled
a large fraction of their mass before reionization {\em and\/} survived
the decay of their orbit as a result of dynamical friction {\em and\/}
avoided tidal disruption by the Milky Way potential.  
As shown in Fig.~1 of BKW,
the number of tidally disrupted objects is similar to the number of surviving
dwarf satellites.
Here we show that the disrupted satellites produce a population of
stars whose total mass and radial profile are consistent with observations
of the Milky Way's stellar halo.  In the outer halo, where the number
of contributing satellites is relatively small and the orbital times
are long, the model predicts substantial substructure, which should
be detectable with the SDSS and other deep, wide-angle surveys.

Our model is simple and is unlikely to be accurate in full
quantitative detail.  However, the qualitative predictions should be
characteristic of conventional CDM cosmologies combined with straightforward
assumptions about the star formation in low-mass dark matter potential
wells.  If the
observed stellar halo is found to be radically different from
these predictions, it will mean that either our star formation assumptions
or the CDM predictions for hierarchical small-scale structure in the
dark matter distribution are incorrect. In this sense, studies of the stellar
distribution in the outer halo can play a valuable role in 
testing more general ideas about galaxy formation.

For convenience, we will focus on predictions for the RR Lyrae
distribution.  RR Lyrae are especially useful probes of the stellar halo
because they are relatively easy to identify,
they are luminous enough
to be detected out to large distances 
($r\sim 100$~kpc), and they are nearly standard candles and
therefore yield 3-dimensional maps.  RR Lyrae
are believed to be good
tracers of the more general halo stellar distribution, and 
they have often been used in kinematic
studies of the halo (Hawkins 1984).  
Finally, while RR Lyrae are numerous enough to trace halo substructure,
they are rare enough that we can construct numerical
realizations that contain every individual RR Lyrae star.  
The stellar density fluctuations predicted by our model are far in excess
of Poisson fluctuations, so it
is straightforward to scale our predictions to other
halo tracers like blue horizontal branch stars or carbon stars, just by
putting in appropriate stellar population weights.

The remainder of the paper is organized as follows. In \S~\ref{sec:method}
we will describe our model and assumptions. Specifically, we will describe
our semi-analytic method for following the accretion and orbital evolution
of satellite galaxies 
in \S~\ref{sec:modelling_accretion} and our modeling of stellar
debris of disrupted satellites in \S~\ref{sec:modelling_debris}. 
We present our results in \S~\ref{sec:results}. We finish with discussion 
and conclusions in \S~\ref{sec:discussion}. 

\bigskip

\section{Method}
\label{sec:method}

\subsection{Accretion and orbital evolution of satellites}
\label{sec:modelling_accretion}

We use a semi-analytic method to trace the accretion
history and orbital evolution of satellite galaxies within a typical
Milky Way-size dark halo.\footnote{In this paper, the term ``halo''
sometimes refers to a dark matter halo and sometimes to a stellar halo.
Usually the meaning is clear from context, but we will specify ``dark''
or ``stellar'' where necessary.  The term ``subhalo'' always refers
to a dark matter halo, one that is accreted into a larger dark matter
halo before redshift zero.}
Detailed description of the model is given
in BKW.  Here, we briefly review its essential aspects. The model uses
the extended Press-Schechter formalism (Bond et al.  1991; Lacey \&
Cole 1993) to construct the accretion history for each
galactic halo. The mass of a halo in this formalism is accumulated via
accretion of individual {\em subhalos} of different masses, and we keep
track of all the accreted subhalos down to some minimum mass.  The
second part of our model is a semi-analytic prescription for orbital
evolution of the accreted subhalos. This prescription is used to
determine whether a subhalo survives to the present day, is tidally
disrupted, or is dragged into the central galaxy by dynamical friction.
Only subhalos that
accreted a significant fraction ($\gsim 0.2-0.3$) of their mass
before intergalactic gas was reheated during the epoch of reionization
are assumed to host luminous galaxies. 
In the model presented in this paper, the stellar halo
of Milky Way-like galaxies is formed from the debris of those
subhalos that once hosted luminous galaxies but were tidally
disrupted before the present day.  
For our analysis, we  adopt a flat CDM   model with a non-zero  vacuum
energy and the following parameters: $\Omega_m = 0.3,
\Omega_{\Lambda} = 0.7, h=0.7, \sigma_8=1.0$, where $\sigma_8$ is the 
rms fluctuation  on the  scale of  $8h^{-1}$ Mpc,   $h$ is the  Hubble 
constant in    units  of $100  \ks {\rm    Mpc}^{-1}$,  and $\Omega_m$  and 
$\Omega_{\Lambda}$ are the density  contributions  of matter and   the
vacuum respectively in units of the critical density.

We assume that the density profile of each dark matter halo is described by the
NFW profile (Navarro, Frenk, \& White 1997): $\rho_{\rm NFW}(x)
\propto x^{-1}(1+x)^{-2}$, where $x=r/r_s$, and $r_{s}$ is a
characteristic inner radius.  Given a halo of mass $\Mvir$ at redshift
$z$, the model of Bullock et al. (2000a) supplies the typical $r_s$
value and specifies the profile completely.  The circular velocity
curve, $v^2(r) \equiv GM(r)/r$, peaks at a value $v_m$ at a radius
$r_{\rm m} \simeq 2.16 r_{\rm s}$.

We use the merger tree method of Somerville \& Kolatt (1999) to
construct mass growth and halo accretion histories for an ensemble of
galaxy-sized dark matter halos.  We start with halos of mass $M_{\rm
  vir}=1.1 \times 10^{12} \hMsun$, at $z=0$, and trace subhalo
accretion histories back to $z=10$.  We record the mass growth for
the primary halo, $M_{\rm vir}(z)$, as well as the mass of each
accreted subhalo, $M_a$, and the time of its accretion, $t_a$ (or
$z_a$).  We assign the subhalo $v_m$ according to the mass-velocity
relation at the epoch of accretion.  For the results presented below,
we use 100 ensembles of formation histories for galactic host halos.
We obtain very similar results if this number is increased.

Each subhalo is assigned an initial orbital circularity, $\epsilon$,
defined as the ratio of the angular momentum of the subhalo to that of
a circular orbit with the same energy, $\epsilon \equiv J/J_{c}$.  We
choose $\epsilon$ randomly in the range $0.1-1.0$ (Ghigna et al. 1998).  To
determine whether the accreted halo's orbit will decay, we use
Chandrasekhar's formula to calculate the decay time, $\tau_{DF}^c$, of
the orbit's circular radius $r_c$ --- the radius of a circular orbit
with the same energy as the actual orbit.  Each subhalo is assumed to
start at a randomly assigned radius $r_c^a = (0.4-0.75) \Rvir(t_a)$,
where $\Rvir(t_a)$ is the virial radius of the host halo at the time
of accretion.  
We determined this distribution of circular radii by
measuring the range of binding energies of subhalos 
in the ART simulations used by Klypin et al (1999a).  
Once $\tau_{\small DF}^{\rm c}$
is known, the decay time for the given circularity is $\tau_{\small
  DF} = \tau_{\small DF}^{\rm c} \epsilon^{0.4}$ (Colpi et al. 1999).
If $\tau_{DF}$ is smaller than the time left between accretion and
$z=0$, $\tau_{DF} \le t_0 - t_a$, then the subhalo will merge with the
central object. In our modeling of the stellar halo, we do not
consider the contribution due to galaxies that subsequently merge with
the central object.  Due to the rapid decay of the orbits, any debris
associated with these objects will likely remained confined within the
radius where stripping first becomes important, typically $r \la
10$~kpc.  For this reason, we consider the predictions for the stellar
distribution only for galactocentric radii $r > 10$~kpc.

If $\tau_{DF}$ is too long for the orbit to have decayed completely
($\tau_{DF} > t_0 - t_a$), we check whether the subhalo would have
been tidally disrupted.  We assume that the halo is disrupted if the
tidal radius becomes smaller than $r_{\rm m}$.  The tidal radius,
$r_t$, is determined at the pericenter of the orbit at $z=0$, where
the tides are the strongest, following Klypin et al. (1999b).  If $r_t
\le r_{\rm max}$ we declare the subhalo to be tidally destroyed and
record its mass and orbital parameters so that we may model the
evolution of its tidal debris (\S 2.2).

The resulting average mass functions for all accreted halos and the
subset of the halos that were tidally disrupted are shown in Figure~1
with the thin dashed and solid lines, respectively.  The error bars
represent the run-to-run dispersion over 100 realizations.  For comparison,
the dotted line
corresponds to halos that were dragged to the halo center as a result of
dynamical friction.  As expected, the dynamical friction is more
efficient for massive subhalos. The mass function of surviving
subhalos is similar to that of the disrupted halos, but we omit it
here to preserve visual clarity.

As in BKW, we assume that of all subhalos with $v_m < 30 \ks$ only
those that accreted a fraction $f$ of their gas before the redshift of
reionization, $\zre$, host luminous galaxies.  Although our results
are consistent with the observed abundance of dwarf satellites for a
range of values of $f$ and $\zre$, here we use the fiducial values
of BKW, $f=0.3$ and $z_{\rm re}=8$.  
We expect that our results for disrupted satellites would be similar
if we chose other $f$, $\zre$ combinations that also match the
observed (surviving) dwarf satellite population.
For a given subhalo of mass $M_a$ and
accretion redshift $z_a$, we use equation (2.26) of LC93 to
probabilistically assign the redshift, $z_f$, at which the main
progenitor of the subhalo reached mass $M_f = f M_a$ for the first
time.  The subhalo hosts a luminous galaxy only if $z_f \ge \zre$.  We
also assume that subhalos with $v_m < v_l = 10 \ks$ do not host
galaxies, since any gas that was initially accreted in these small
systems would be unable to cool, and it should quickly boil out of the
halo after reionization (e.g., Barkana \& Loeb 1999).  Our results do
not change if we vary $v_l$ by $50\%$.  The thick solid line in
Figure~1 shows the average mass function of disrupted halos that once hosted
luminous galaxies.  The mass function of the surviving galaxies is similar
to that of the disrupted galaxies, but

{\pspicture(0.5,4.9)(12.0,19.) 
\rput[tl]{0}(-0.2,19.0){\epsfxsize=8.9cm 
\epsffile{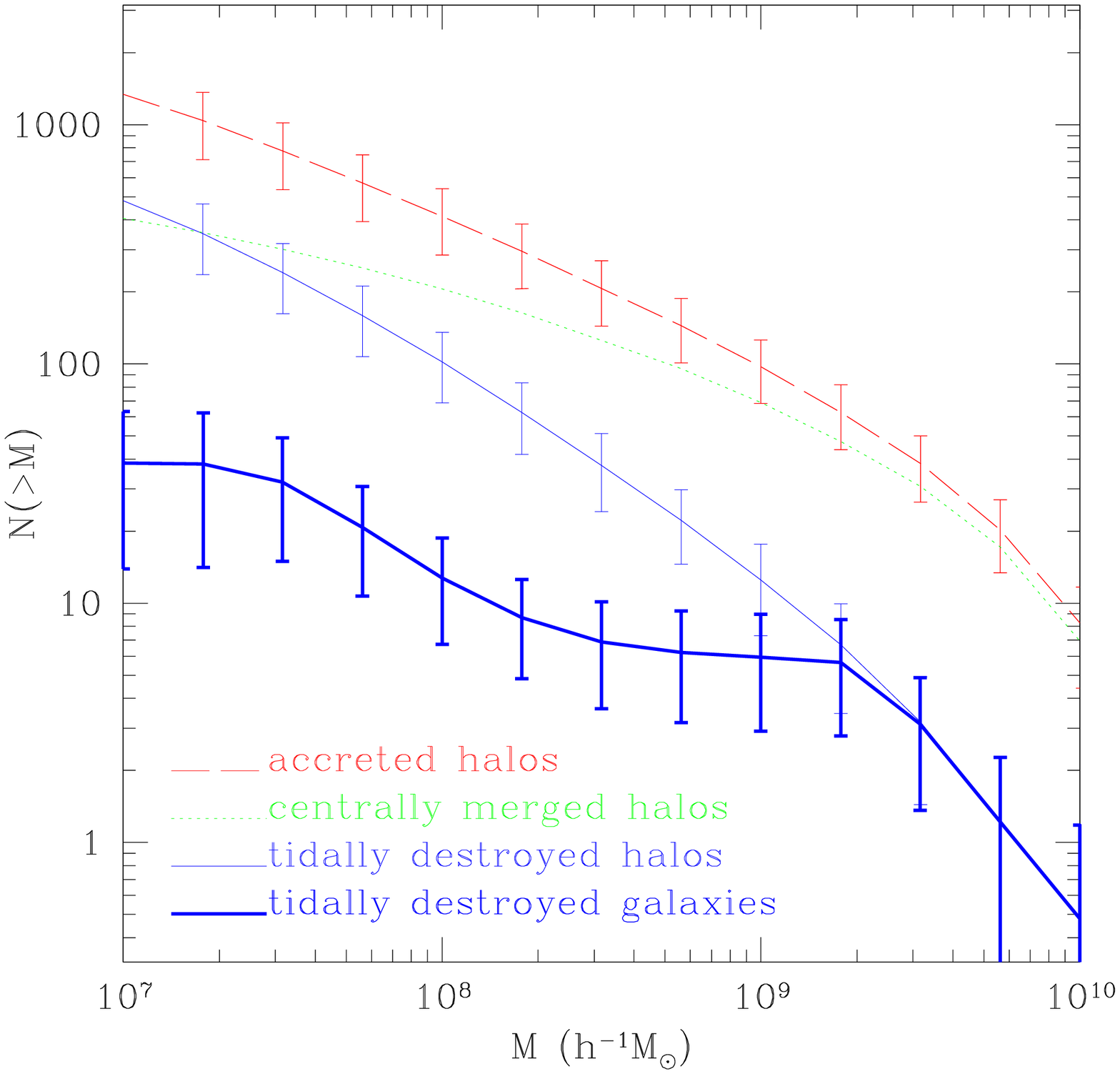}} 
\rput[tl]{0}(.1,10){ 
\begin{minipage}{8.7cm} 
  \small\parindent=4.5mm {\sc Fig.}~1.--- Cumulative mass function
  of  all accreted dark  matter subhalos ({\em dashed line}), 
  the fraction that decayed due to dynamical friction ({\em dotted line}),
  tidally disrupted halos ({\em thin solid line}), and the fraction
  of disrupted halos that host galaxies ({\em thick solid line}).
  Not shown is the mass function of surviving halos, which
  is similar to that of disrupted halos, and the mass function of
  surviving {\em galaxies}, which is roughly half that of 
  the disrupted galaxies (reflecting the tendency for surviving halos
  to have been accreted later).
   The  mass function represents  the average
  over    100  merger  histories    for  host  halos   of mass     $M_{\rm
  vir}(z=0)=1.1\times 10^{12}h^{-1}{\ }{\rm M_{\odot}}$. The errorbars
  show   the dispersion over the different   merger histories.  
\end{minipage} 
 } 
\endpspicture} 

\noindent
lower in amplitude by about a
factor of two.  Surviving halos are typically accreted later than
tidally destroyed halos, and 
they are less likely to form before reionization and host a galaxy.

\subsection{Modeling stellar tidal debris}
\label{sec:modelling_debris}

Estimating the number of tidally stripped stars, and RR Lyrae stars
in particular, requires several uncertain assumptions about gas
cooling, star formation, and stellar population morphology.  The
following approximations are extremely simplified, and the estimated
number of disrupted stars for each object is uncertain at the factor
of $2-3$ level. This uncertainty is passed on to the overall amplitude
of calculated stellar density distribution in the halo. Nevertheless,
our statistical measures of substructure depend only on the ratio to
the background density, and they are thus largely insensitive to the precise
values we assume.  Furthermore, we adopt the same assumptions 
that we used in BKW to obtain consistency with the observed dwarf
satellite population, and this matching normalizes out some of the
uncertainties in the overall stellar halo amplitude.

We estimate the luminosity of every disrupted luminous galaxy assuming
that it has a baryonic mass 
$fM_a(\Omega_b/\Omega_m)$ (the mass of
baryons accreted at $z>\zre$). We assume that a fraction $\epsilon_*$ of
this baryonic mass
is converted to a stellar population with mass-to-light ratio $M_*/L_V$.
The total mass to light ratio of the subhalo is thus 
\beq \left(\frac{M_{\rm vir}}{L_V}\right) = f^{-1}
\left(\frac{\Omega_m}{\Omega_b}\right) \left(\frac{M_*}{L_V}\right)
\epsilon_*^{-1}.  
\eeq 
Adopting $M_*/L_V \simeq 0.7$, typical for
galactic disk stars (e.g., Binney \& Merrifield 1998), $\Omega_m/\Omega_b
\simeq 7$ (based on $\Omega_m=0.3,$ $h=0.7$, and $\Omega_b h^2 \simeq
0.02$ from Burles \& Tytler 1998), and $\epsilon_* = 0.5$, we obtain
$(M/L_V) \simeq 10 f^{-1} \simeq 33$.  We estimate the number of
horizontal branch stars in each galaxy using $N_{HB} = L_V/(540
L_{\odot})$ (Preston, Shectman, \& Beers 1991).  The fraction of the
horizontal branch stars that are RR Lyrae variables is strongly dependent on 
the metallicities and ages of the populations, and it will vary significantly
from one object to another.  For simplicity, we assume that all
disrupted objects have $N_{RR} = 0.3 N_{HB}$.  This fraction is high
compared to local halo stars, but it is consistent with fractions
observed for more distant ($r \ga 10$~kpc) globular clusters, likely
reflecting the tendency for the outer halo to be younger (Preston et al.
1991; Brocato et al. 1996; Layden 1998).

For each disrupted luminous subhalo, we randomly assign a direction
for its angular momentum vector.  This direction fixes the plane
of the orbit since we assume that the dark halo
potential is spherical (i.e., no orbital precession).
We follow the orbit from the time it was accreted at $t_a$ to $t_0$
using \beqa \frac{dr}{dt} & = & \pm v(r_c) \sqrt{ \frac{2}{v^2(r_c)}
  [\Phi(r_e) - \Phi(r)] + 1 -  \frac{\epsilon^2 r_c^2}{r^2}}, \\
\frac{d \psi}{dt} & = & \frac{v(r_c) r_c \epsilon}{r^2}, \eeqa where
$r$ is the distance to the galactic center, $\psi$ is the angle in the
orbital plane, $\Phi(r) = - 4.6 v_m^2 \ln(1+x)/x$ is the potential of
the host, and the $\pm$ sign signifies whether the object is
approaching its apocenter or pericenter, respectively.  We will work
in the approximation that the dynamical friction timescale is long
compared to the time remaining for the orbital evolution and
disruption: $\tau_{DF} \gg t_0 - t_a$.  This is a good approximation
for $\sim 90\%$ of the disrupted halos --- not surprisingly, since we
have deliberately restricted our analysis to halos whose orbits have
not decayed (i.e., long $\tau_{DF}$).  In order to approximately
account for cases where this approximation breaks down, we start
integrating the satellite orbit at $t=t_a$ but set its starting
radius equal to the circular radius it will have decayed to by
$t=t_0$: $r(t_a)=r_c(t_0) \equiv r_c$.  The initial value for the
angle, $\psi(t_a)$, is chosen randomly, and we assume that the
satellite is initially infalling (approaching the pericenter).

We assume that the satellite is tidally disrupted after the first
passage of its orbit pericenter.  At this time, the tidal debris will
obtain an energy distribution from the encounter.  Our model for the
evolution of the debris along the tidal tail is motivated by numerical 
results of Johnston (1998), who showed that the following approximations
provide good description of the positions of stripped
particles in her simulations.
The typical energy scale of  the debris is set  by  the change in  the
host halo potential  energy  over the  size of  the tidally  disrupted
object,
\beqa
        \varepsilon = r_t \frac{d \Phi}{d r} \simeq v_m^2 
                             \left(\frac{r_t}{r_p}\right),
\eeqa
where $r_p$ is the pericenter radius of the orbit. 
The last approximation is exactly  true for a logarithmic potential (a
singular isothermal density profile).  We assume that the satellite is
completely  disrupted after  the  first passage, and  that the energy of
the debris is evenly distributed over the  energy range $-\varepsilon > dE
> \varepsilon$ (Evans \& Kochanek 1989).  This assumption ignores the
possibility of disruption over several orbits, but we find that our
results are  robust to the choice of  distribution and do  not change
significantly  if  the energy range is altered by  $50  \%$.  Using intuition
gained  from a circular  orbit  within  a singular isothermal  density
background, we may  estimate how a  change in energy  from $E$ to $E +
dE$  affects the orbit   of a particle.    In this approximation,  the
azimuthal period, $T_{\psi}$,  and radial period, $T_{r}$, depend only on
the orbit energy,  and they are both increased or decreased  depending on the
sign of the deposited energy: $T_{\psi,r}(E+dE) = \tau T_{\psi,r}(E)$,
where
\beqa
 \tau = \exp\left(\frac{dE}{v_m^2}\right).
\eeqa 
This result  allows us to map   the orbital trajectory of  the initial
object  with energy $E$ (Eqs.~4 and 5) to  that  of a debris particle
with     energy    $E+dE$     via    [$r(t)$,$\psi(t)$]  $\rightarrow$
[$r(t/\tau)$,$\psi(t/\tau)$].

For each RR Lyrae star in the disrupted galaxy, we  assign a change in
energy  $dE$ and integrate  the  orbital  equations to determine  its
position at  $z=0$.  Since the disrupted  galaxy will have some finite
spherical  extent,  we add a random offset to
this calculated central orbit position.  The magnitude of this offset is
a Gaussian  deviate with  dispersion $2$~kpc, the typical
optical radius for a dwarf galaxy (Mateo 1998), and the direction 
is random.

\section{Results}
\label{sec:results}

Figures  2 and 3  show two realizations  of  the RR Lyrae distribution
from disrupted satellites  in sky-projected galactic coordinates.  The
panels in  each  figure correspond to the    indicated radial bins  in
galacto-centric radius.  Qualitatively, it is clear that substructures
become more  pronounced at larger radii. 
This radial trend reflects the smaller number of disrupted satellites
with large apocenters and the longer periods of their orbits, which
reduces the extent of debris spreading along the orbit.
A comparison of the maps in Figures 2 and 3
illustrates the differences between different merger histories of the host
halo. One can see that  the stochastic variations in merger history 
at fixed host mass lead to substantial variations in the appearance
and abundance of substructures.

In light  of the SDSS results  referred to in \S~1, the most interesting
predictions of the model are the radial number density profile of halo
stars and the clumpiness and spatial  extent of the stellar distribution.  
Figure~4 shows   radial number density profiles of   the halo RR Lyrae
stars.  The long  dashed lines in  each panel represent the  power law
($n_{\ast}\propto r^{-3}$)  RR    Lyrae density profile  derived    by
Wetterer  \&  McGraw (1996), based  on  their  large  compilation of RR
Lyrae.  The solid  points show the  profile computed

\pspicture(0,0)(18.5,19.5)

\rput[tl]{0}(2.,19.){\epsfxsize=15cm
\epsffile{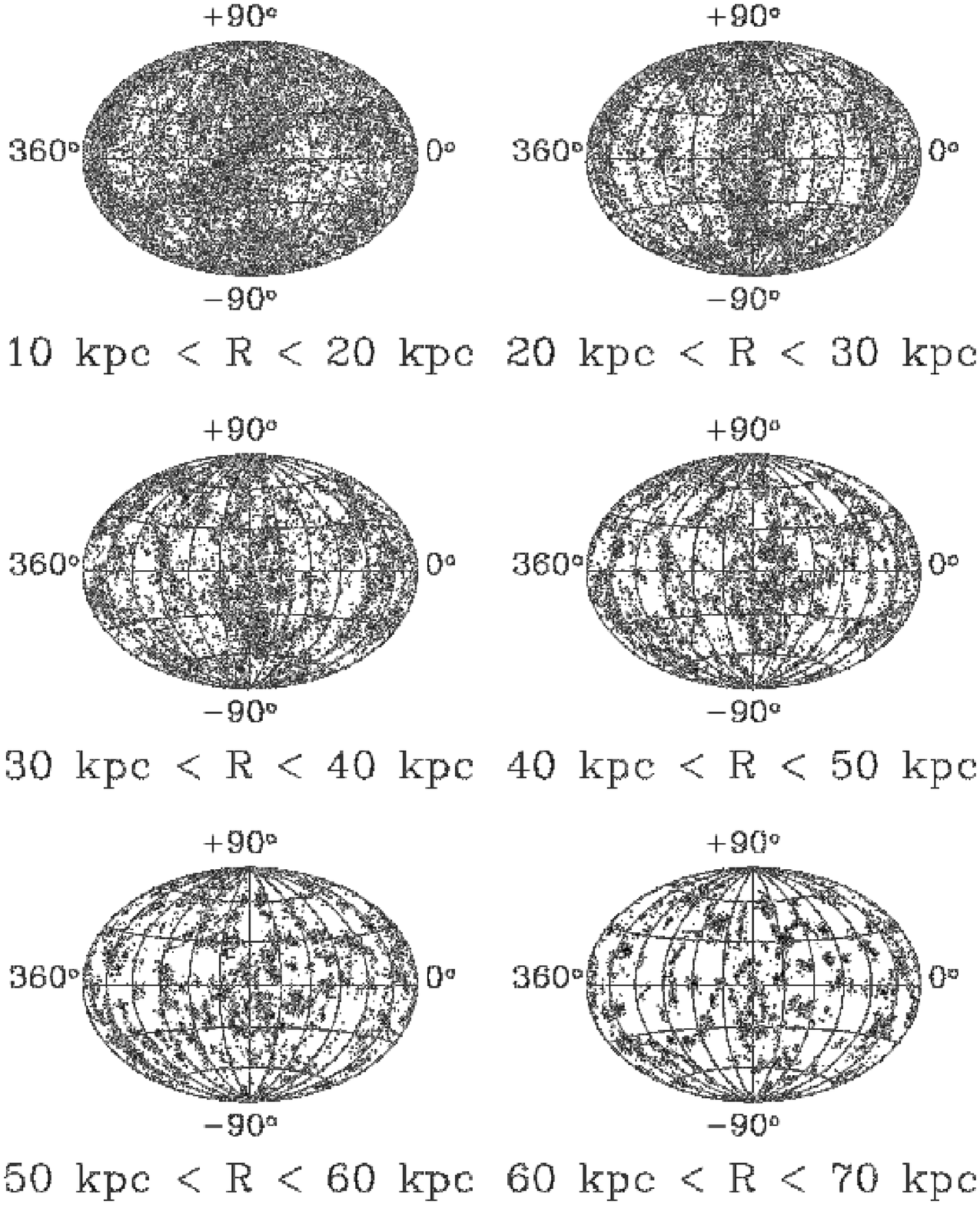}}

\rput[tl]{0}(0.,0.){
\begin{minipage}{18.5cm}
  \small\parindent=3.5mm {\sc Fig.}~2.---  Distribution of stripped
  stars in various radial bins projected on the sky.  Each point
  represents an RR Lyrae star, and the number of stars in each radial
  bin, starting in the upper left panel, is 11331, 9052, 8237,
  7182, 6173, and 5076.  These views are centered on the Galactic Center,
  but shifting to a solar origin makes no qualitative difference.
\end{minipage} 
 } 
\endpspicture

\newpage

\pspicture(0,0)(18.5,19.5)
\rput[tl]{0}(2.,19.){\epsfxsize=15cm
\epsffile{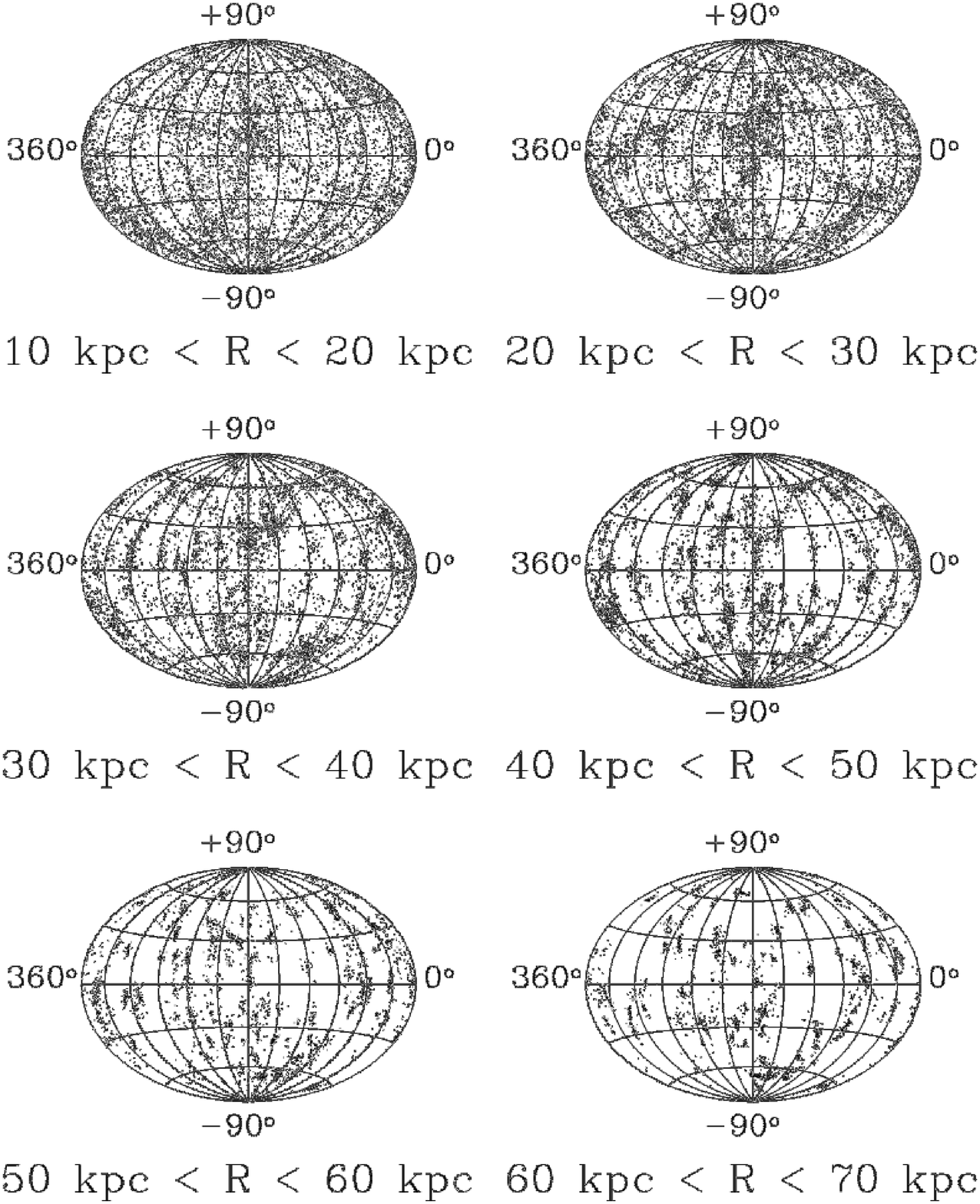}}

\rput[tl]{0}(0.,0.){
\begin{minipage}{18.5cm}
  \small\parindent=3.5mm {\sc Fig.}~3.--- Same as Figure~2, but
  for a different merger history realization.  Each point represents
  an RR Lyrae, and the number of stars in each radial bin, starting in
  the the upper left panel, is 6511, 6402, 4962, 4363. 2783, and 2077.
\end{minipage} 
 } 
\endpspicture

\newpage

{\pspicture(0.5,4.9)(12.0,18.) 
\rput[tl]{0}(0.2,18.4){\epsfxsize=9.cm 
\epsffile{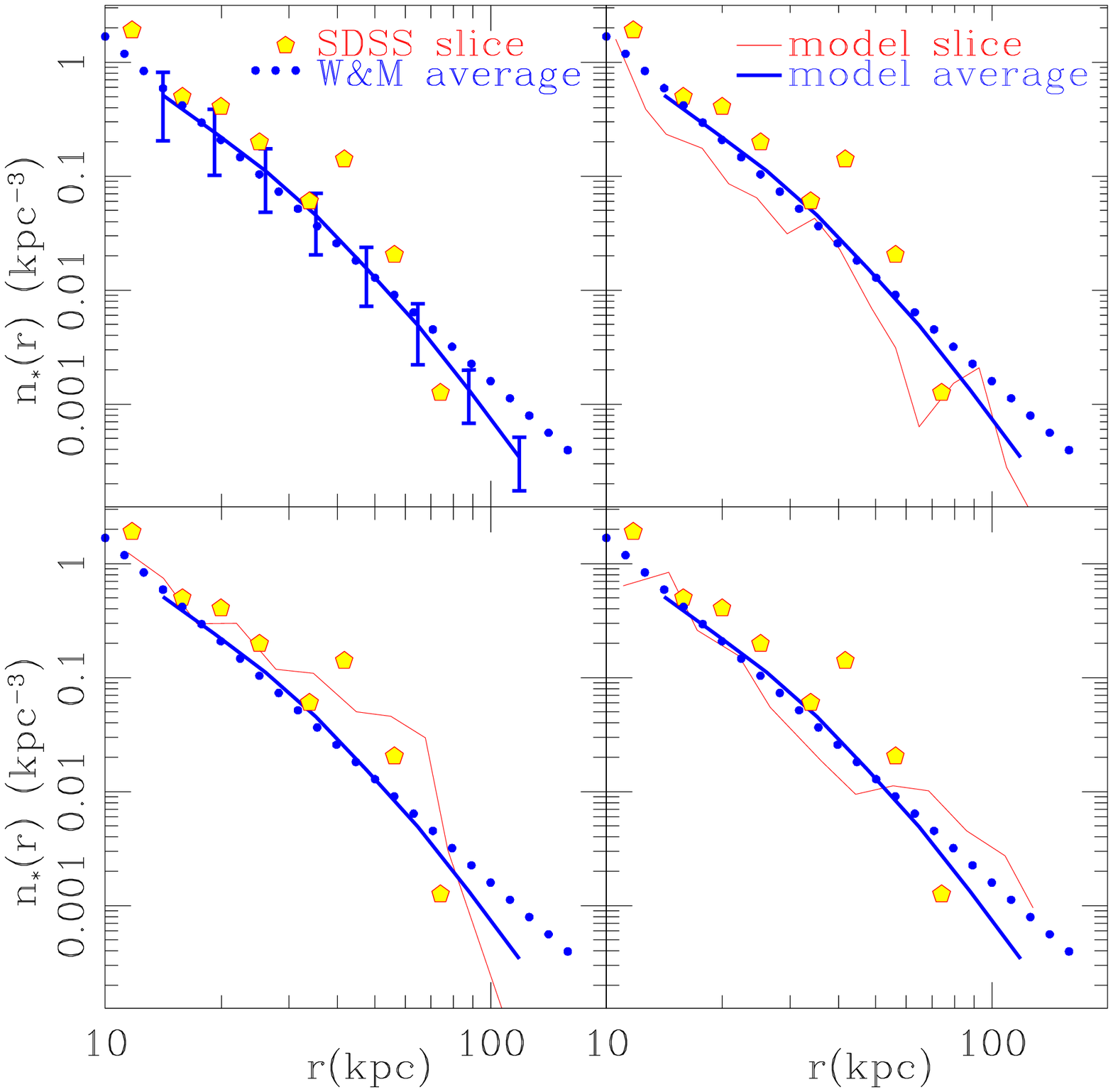}} 
\rput[tl]{0}(.1,9){ 
\begin{minipage}{8.7cm} 
  \small\parindent=4.5mm {\sc Fig.}~4.--- The average RR Lyrae density
  profile (thick solid line) compared with the Ivezi\'{c} et al.
  (2000) SDSS data (solid points) and the power law determined by
  Wetterer \& McGraw (1996) (dashed line).  In the upper left panel,
  the error bars reflect the dispersion in the average from
  realization to realization.  The thin solid lines in the other three
  panels are results of random strips similar in solid angle and geometry
  to the strips used to obtain the SDSS measurements ($\sim 1$ degree wide,
  $100$ square degree strip).
\end{minipage} 
 } 
\endpspicture} 

\noindent 
by Ivezi\'{c} et
al. (2000) for  their sample of RR Lyrae  candidates obtained from  
SDSS  commissioning  data, which covers roughly a one-degree wide,
100 square degree strip 
of sky. Note that  the  SDSS and  Wetterer \& McGraw
profiles agree well at $\lsim  35$~kpc. At  larger radii, however, the
 SDSS sample  shows two significant  deviations from the  smooth
$n_{\ast}\propto  r^{-3}$ profile: a  ``bump''  in number density at
$r\approx 40$~kpc and a sharp drop at $r\gsim  50$~kpc.  As noted by
Ivezi\'{c} et al., this structure in the radial profile
likely indicates significant clumpiness of the
stellar halo at these galactocentric radii, and the bump in
particular is associated with an identifiable coherent structure
containing $\sim 70$ RR Lyrae within the observed region.

The  model predictions are shown in  Figure~4 by  thick and thin solid
lines.  The thick solid lines represent  the  computed RR Lyrae number
density profile averaged over all merging history realizations and the
full sky.  The error bars in the  upper left panel show the dispersion
from realization to   realization around  this  average, demonstrating
that stochastic variations in merger histories lead  to a 
factor of $\sim 2$ rms  variation in 
the  overall normalization  of the predicted halo
density profile.  In the remaining  three panels, the thin solid lines
show examples of  density  profiles derived from   a single host  halo
realization,  viewed through three  randomly  chosen strips similar in
solid angle and geometry to the strips used to derive the SDSS sample.

{\pspicture(0.5,4.9)(12.0,17.3) 
\rput[tl]{0}(0.2,17.2){\epsfxsize=9.cm 
\epsffile{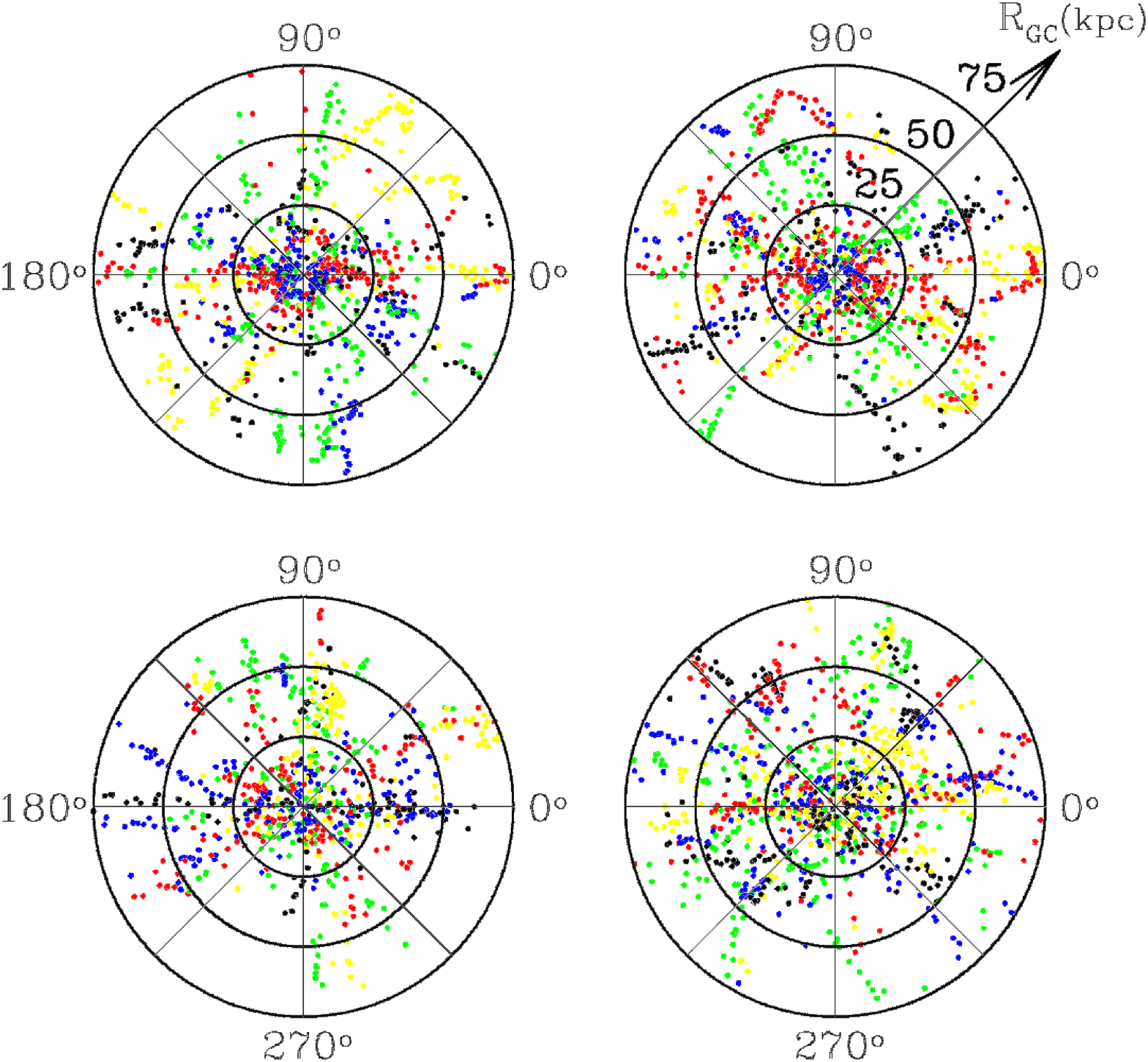}} 
\rput[tl]{0}(.1,9){ 
\begin{minipage}{8.7cm} 
  \small\parindent=4.5mm {\sc Fig.}~5.--- The radial distributions of
  disrupted RR Lyrae stars in 1 degree wide, great circle slices through
  a single model halo realization.  This halo formed from about
  60 disrupted satellites.
  Shown are four random cuts of great circle planes through 
  the halo center.
  Each point
  represents a single RR Lyrae star. 
  Concentric circles indicate galactocentric radii of 25, 50, and 75~kpc.
  Note that apparent ``clumpiness'' of the 
stellar distribution increases with increasing radius.   
\end{minipage} 
 } 
\endpspicture} 
\label{fig:grcslices1}

The first remarkable feature of Figure~4 is the agreement of the
predicted average profile with the slope and amplitude found by 
Wetterer \& McGraw (1996).  As discussed in \S 2.2, the predicted
amplitude is uncertain by a factor of several, and even the statistical
fluctuations from one galaxy to another are significant, so the
degree of agreement must be somewhat fortuitous.
However, it is worth noting that we did not adjust any parameters
to fit the observed halo profile but chose ``best guess'' values based
on other considerations --- in particular, the requirement of matching
the observed dwarf satellite population.
Figure~4 suggests that disruption of accreted satellites can
produce not just substructure in the stellar halo but the entire
stellar halo itself.  If we have overestimated the number of RR Lyrae
per unit dark matter mass, then there is room for another physical
mechanism that creates a smooth underlying halo, but it seems
that no such additional mechanism is necessary.

The second remarkable feature of Figure~4 is the jaggedness of
the observed and predicted profiles of individual strips, which
becomes especially pronounced at radii $r \ga 40$~kpc.
These large fluctuations reflect the substructure 
visible
in Figures~2 and~3.  The steep drop in the SDSS counts between
60 and 70~kpc suggests detection of an ``edge''
 of the stellar
halo (Ivezic et al.\ 2000).  However, the second of our numerical
realizations shows an equally sharp edge, even though the average
halo profile is smooth.  Our model predicts a gradual steepening of the
halo profile at $r \ga 60$~kpc, but although surveys in small solid
angles should show large count fluctuations, the profile averaged

{\pspicture(0.5,4.9)(12.0,16.9) 
\rput[tl]{0}(0.2,17.2){\epsfxsize=9.cm 
\epsffile{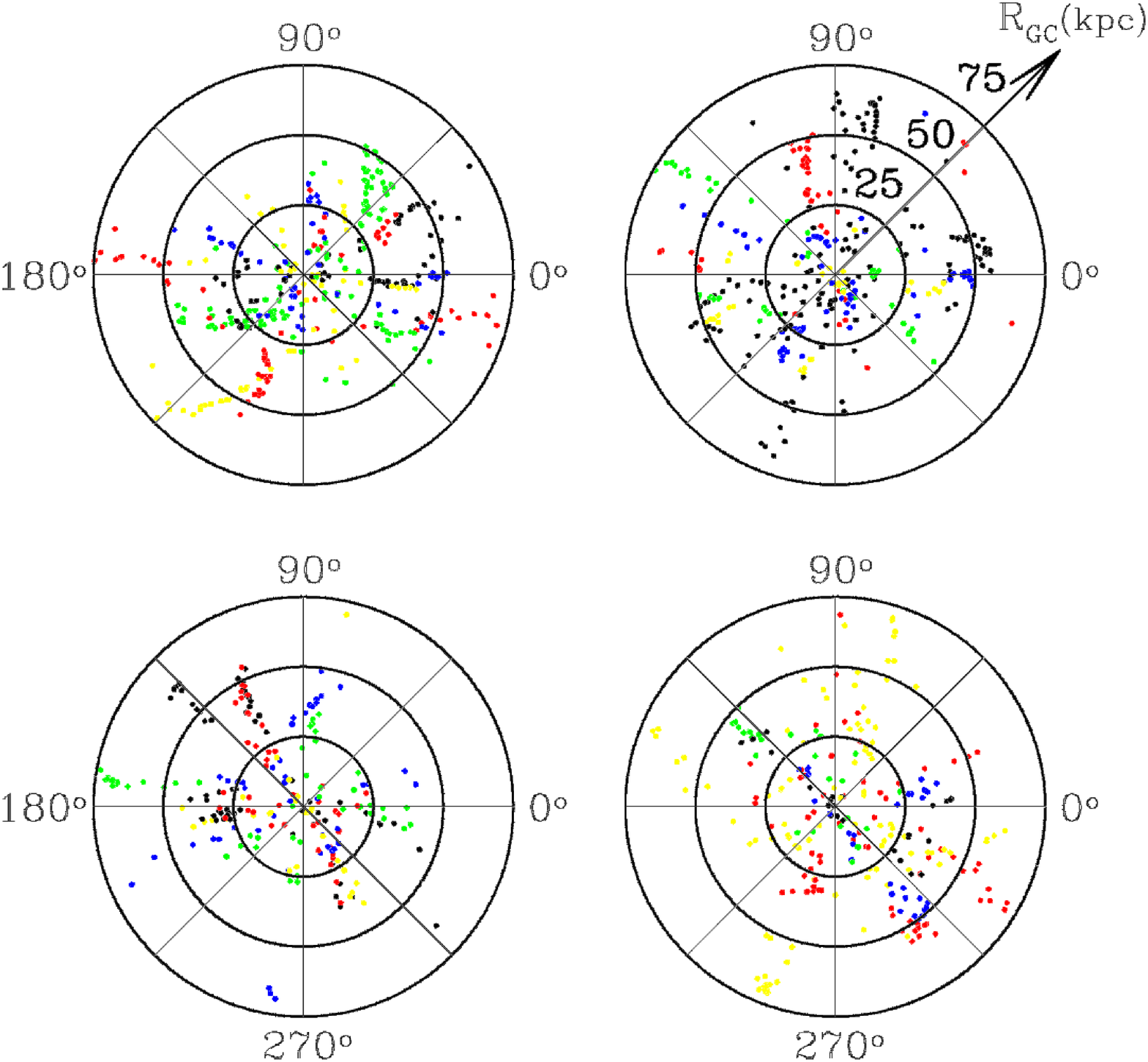}} 
\rput[tl]{0}(.1,7){ 
\begin{minipage}{8.7cm} 
  \small\parindent=4.5mm {\sc Fig.}~6.--- Same as 
Figure~5, but for 
a different merger history realization of the host halo.   The host
in this realization accreted about 20 luminous satellites over 
its history. 

\end{minipage} 
 } 
\endpspicture} 
\label{fig:grcslices2}

\noindent
over the full sky should not cut off sharply.

Figures~5 and~6 present a  different view of the  structure associated
with disrupted satellites,  in a form more  comparable to the plots of
Yanny et  al.\ (2000).  Each figure  shows a one degree wide, randomly
oriented,  great circle slice  through  a realization  of the RR Lyrae
distribution.  The two figures  show stellar halos for different Monte
Carlo accretion   histories, one with  a total  of about  $60$ tidally
destroyed galaxies (Figure~5) and one  with a more quiescent accretion
history  and about $20$  tidally destroyed galaxies  (Figure 6).  This
range roughly covers  the scatter in  the number of  disruption events
expected from galaxy to galaxy.  The three concentric circles indicate
galactocentric  radii   of 25,  50, and   75~kpc. In   both figures,
structures  associated with  the  individual  disrupted objects become
more easily identifiable at larger radii.

To  make  our  predictions more  quantitative, we   present two simple
statistical measures of   the  halo clumpiness.    Figure~7  shows the
probability distribution of model RR Lyrae counts in solid angle cells
of different sizes  and for different  ranges of galactocentric radii.
The cells are roughly square on the sky;  they are defined by dividing
the sky map for each  realization into patches of  a  given size.   In
order to  take  out the uncertainty in  the  overall amplitude of  the
density distribution, and to factor  out the variation in the  overall
amplitude  from   one halo  realization  to  another,  the  
counts are
presented in units of the average expected
 number, $\langle N\rangle$,
of  RR Lyrae   in patches of   the  chosen size for  each realization.
Figure~7 shows that the amplitude of patch-to-patch count fluctuations
is   higher for smaller  solid  angles  and for larger  galactocentric
radii.

{\pspicture(0.5,4.9)(12.0,18.) 
\rput[tl]{0}(0.2,18.4){\epsfxsize=9.cm 
\epsffile{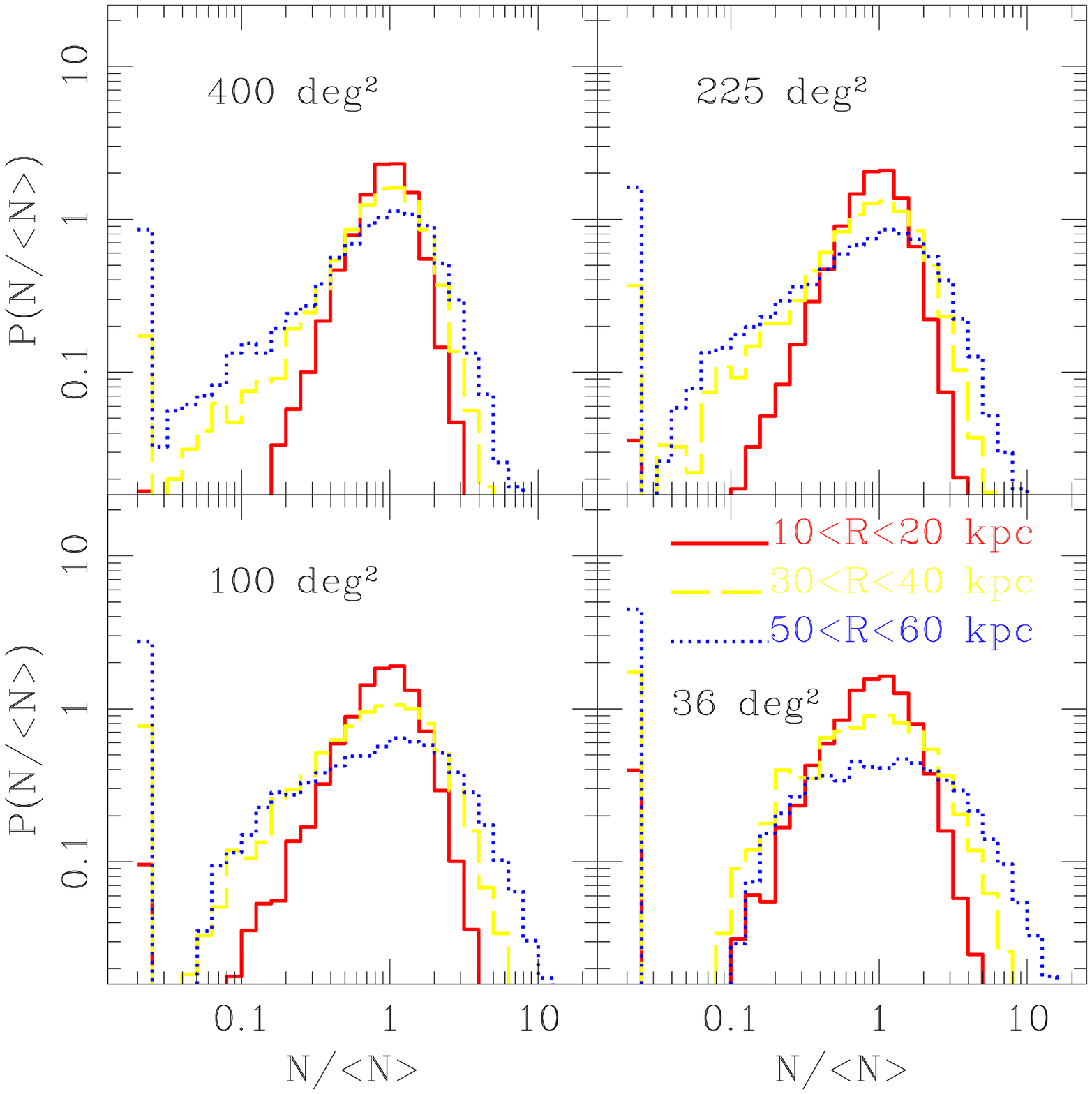}} 
\rput[tl]{0}(.1,9){ 
\begin{minipage}{8.7cm} 
  \small\parindent=4.5mm {\sc Fig.}~7.--- Probability distribution of
  simulated RR Lyrae counts in solid angle cells for various radial
  ranges. Here, $N/\langle N \rangle$ is the measured number of stars within the
  given radial bin and solid angle cell divided by the average number
  for a cell of that size. The cells are roughly square in angle
  size and were defined by dividing the sky of each realization
  into patches of the indicated size, with the observer at the center
  of the halo.  The spikes at small $N/\langle N \rangle$ represent empty cells.
  Note that higher amplitude fluctuations from patch to patch are more
  likely for smaller patch areas and for larger galactocentric radii.
\end{minipage} 
 } 
\endpspicture} 
\label{fig:pd}

Figure~8  shows the rms dispersion    in RR Lyrae counts,   
$\sigma(N/\langle N \rangle)$,  as   a  function of galactocentric    radius  for cells of
different solid angle and geometry.  The solid points/line show counts
in square cells, while open points  represent the counts in one degree
wide strips.  The  error bars for  the solid  points show the standard
deviation of the dispersion for different merger history realizations;
the error bars are similar for the open points.
These error bars reinforce the point made in our illustrative figures
above, namely that the degree of surviving substructure is quite
variable from one realization of the stellar halo to another.The  dashed 
lines show the expected
amplitude of Poisson fluctuations based on the average number of stars
expected within the given solid angle and radial bin. Note that 
the predicted fluctuations are always larger than Poisson fluctuations,
especially at large radii, because they are dominated by fluctuations
in the number of debris streams rather than $\sqrt{N}$ fluctuations
in the number of RR Lyrae.
Again,  the fluctuations are larger  for cells of  smaller solid angle
and for larger  radii.   They are   also larger  for
square patches than for narrow strips of  the  same solid
angle, reflecting the fact that a narrow strip is less likely
to enclose an entire disrupted object and instead encloses fragments
of multiple debris streams.

{\pspicture(0.5,4.9)(12.0,18.) 
\rput[tl]{0}(0.2,18.4){\epsfxsize=9.cm 
\epsffile{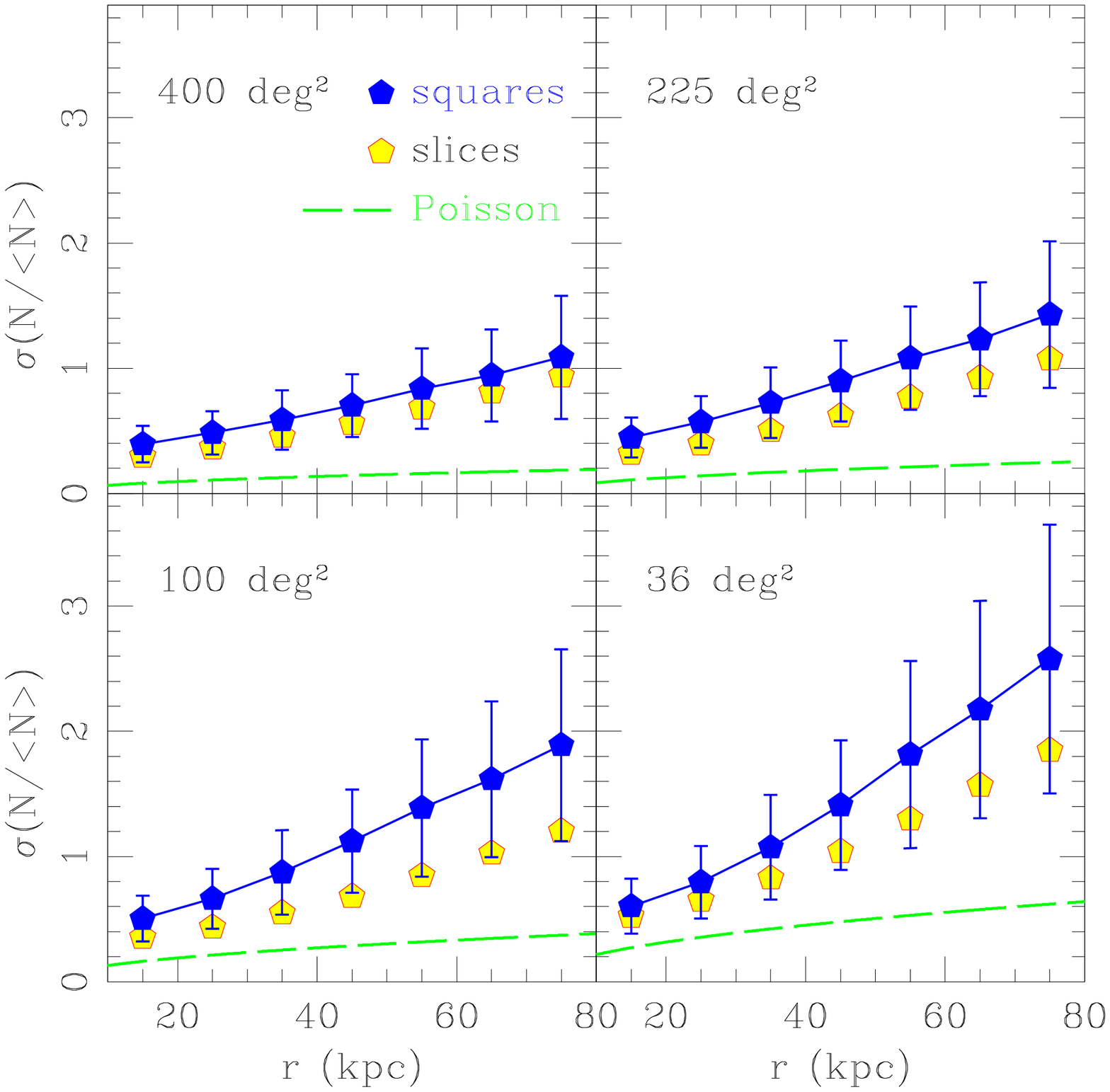}} 
\rput[tl]{0}(.1,9){ 
\begin{minipage}{8.7cm} 
  \small\parindent=4.5mm {\sc Fig.}~8.--- Fractional rms fluctuation 
  of stellar counts
  as a function of radius for sky patches of different solid angle.
  The solid points are for a roughly square angular patch geometry, and
  the open points correspond to one degree wide strips.  The error
  bars represent the dispersion in the rms fluctuation from realization to
  realization for an ensemble of merger histories.  The dashed lines
  show the expected Poisson uncertainty based on the average number of
  RR Lyrae expected within the given solid angle and radial bin.
\end{minipage} 
 } 
\endpspicture} 
\label{fig:variance}

\section{Discussion and conclusions}
\label{sec:discussion}

We have presented a model in which stellar halos of Milky Way 
type galaxies are built via the accretion and tidal disruption 
of satellite galaxies.  The model is based on 
the CDM structure formation scenario, with the crucial assumption 
that only a fraction of  dark matter halos with
circular velocities $\lsim 30{\rm km\ s^{-1}}$ are luminous and host a
sizable stellar system\footnote{All DM halos with circular velocities
  $>30{\rm km\ s^{-1}}$ are assumed to host a luminous galaxy.}. These
luminous halos are those that collapsed and accreted a substantial
fraction of their mass prior to the epoch of reionization, $\zre$.
Only the  accretion and tidal disruption of {\em luminous} dwarf 
galaxies contribute to the build up of the stellar halo in our model.

The model  predicts  an average  density profile for the
stellar halo of $n_{\ast}\propto  r^{-\alpha}$, with $\alpha  \sim 3$
in the range $r\simeq 10-50$~kpc, in good agreement with observations.
The density profiles  of individual realizations of the
host halo  merger  histories, however, exhibit a   substantial scatter
around this average,  with the power law  varying from $\alpha \sim 2.5  -
3.5$.  The reason why the stellar halo density profile is steeper
than that of the dark matter ($\alpha_{DM} \sim 2$, which is
indeed typical of that of the {\em surviving} subhalos in 
our model), is that central satellites are more likely to be 
destroyed than those at large radii, since satellites far from the
galactic center must have extremely eccentric orbits in order to 
pass close enough to be tidally destroyed.
The average predicted profile falls off more steeply at 
$r \ga 50$~kpc, but it does not have a sharp outer boundary.

The amplitude of the mean profile depends on several uncertain factors,
such as the mean stellar luminosity to dark mass ratio of the disrupted
dwarfs and the mean number of RR Lyrae per solar luminosity.
There are also significant (factor of two) statistical fluctuations
in the amplitude from one realization to another.  The amplitude of
the predicted RR Lyrae profile is therefore uncertain by a factor
of a few.  Nonetheless, with ``best guess'' parameter values chosen
on the basis of other considerations, the mean RR Lyrae profile agrees
very well with that determined by Wetterer \& McGraw (1996), in slope
and amplitude.  It therefore appears that disruption of dwarf
satellites is a plausible mechanism for producing the entire stellar
halo within the CDM framework, though it could also co-exist with
some other mechanism.  The uncertainty in our predicted normalization
is reduced by the fact that we require the model to self-consistently
reproduce the velocity function of observed dwarf satellites, a point
that we will return to shortly.

The  main qualitative prediction  of the  model   is the  presence of
significant clumpiness  in the outer  regions ($\gsim 30$  kpc) of the
Galaxy's stellar halo.  This clumpiness is due  to the surviving tidal
debris  of dozens of satellite galaxies  disrupted during evolution of
their  host.  In the inner regions of  the stellar halo ($r\lsim 30$ kpc),
the density distribution is relatively smooth.
At larger radii, however, the clumpiness of
the stellar halo manifests itself when viewed through fixed solid
angle patches in the sky.  For typical modeled stellar halos, 
RR Lyrae profiles of the type observed by Ivezi\'{c} et al.  (2000)
in the  SDSS  commissioning  data (with a coherent structure
at $r \sim 50$ kpc) 
are not uncommon. 

We have quantified our predictions by presenting
some statistical  measures of the ``clumpiness'' in our modeled 
stellar halos.     First, we measured the probability   distribution (see
Fig.~7) of  RR Lyrae  counts,  $N/\langle N\rangle$, in solid
angle cells of different sizes and  for different galactocentric radial
bins (here $\langle N\rangle$ is  the average number of stars expected
in   a cell).   We  also  presented  the rms  width of   this probability
distribution as a function  of galactocentric  radius for solid  angle
cells  of different geometries   (Fig.~8).  These statistics show that
the variance in the stellar counts (i.e., clumpiness  of the stellar halo)
increases with   increasing galactocentric radius and  decreasing solid
angle.  Although current observational  data sets are not sufficiently
large to derive similar  statistics for comparison, future samples  of
RR Lyrae and  A-stars from the SDSS  and 2dF surveys  should make such
comparisons possible.  For cell sizes of 36 deg$^2$ or larger, predicted
fluctuations in RR Lyrae counts are much larger than Poisson fluctuations,
at all galactocentric radii considered.

These    quantitative predictions  depend     on  some  of  the  model
assumptions.  For   instance, one of  the  implicit assumptions in our
analysis is that the dark  matter distribution in the  host halo is nearly
spherically symmetric.  In particular, we do  not include the possible
precession   of satellite orbits. In   the  case of the  significantly
oblate halo,  such precession can  erase signatures of
the  tidal tails   (at least  in  configuration  space), so it would
reduce the predicted clumpiness of the outer halo (e.g., Ibata
et al. 2000a).  
However, the observed narrowness  of the tidal tail of the
Sagittarius dwarf galaxy implies  a nearly spherical halo potential at
$r\lsim  60$   kpc  (Ibata et  al.     2000a).  A nearly spherical  mass
distribution in the  inner    region of a {\em galaxy-mass}   halo  is
consistent with predictions of CDM models (Bullock et  al. 2000b).  We
have  also  neglected  the  effects   of the  disk  component on the
background  potential.  Including this non-spherical central component
would  induce precession in the tidal  orbits (Helmi  \& White 1999)
and smear out residual structure  at small galactic  radii.    But since this
effect would only be important   in  the central  regions ($r \lsim
20$ kpc), the net effect would be to  strengthen the trend of increased
variance in star counts with radius.

The assumption that affects our predictions the most, however, 
is that only a small fraction of
disrupted subhalos host a stellar system and thereby contribute to the
build up of the stellar halo.  In other words, the predicted
properties of the stellar halo depend crucially on the way that we have
solved the dwarf satellite problem.  This problem, namely
that the predicted number of dark halos with  circular
velocities  $\lsim 30  \ks$ is much larger than the 
observed number of dwarf satellites within the virialized dark  halos  of the
Milky Way and M31 (Kauffmann et al.  1993;  Klypin et al.  1999a; Moore
et al. 1999), is one of the few outstanding problems of the 
conventional inflationary CDM model, which, with the inclusion of a
cosmological constant, accounts well for a wide variety of other 
observational data.

Other proposed solutions to the dwarf satellite problem include modifying
the inflationary fluctuation spectrum (Kamionkowski \& Liddle 2000)
or modifying the properties of dark matter by making it warm
(WDM; e.g., Hogan \& Dalcanton 2000) or 
self-interacting (SIDM; e.g., Spergel \& Steinhardt  2000).
In the SIDM model, halos collapse and accrete mass in a similar manner
to conventional CDM halos.  However, the number of surviving subhalos
within a Milky Way mass halo is reduced because lower concentration
of the SIDM halos makes them easier to disrupt and because interactions
lead to ``ram pressure'' stripping of the dark halos.
Relative to our reionization solution, the SIDM solution 
 seems to predict a more massive and more extended stellar halo built from
a larger number of tidal streams, since the abundance of dwarf satellites
is reduced by a higher efficiency of satellite disruption rather
than suppression of star formation in low-mass systems.
Indeed, the SIDM model seems at some risk of overproducing the stellar
halo.  However, if the predicted stellar halo were normalized to the
mean stellar density of the observed one, the SIDM model would probably
predict less substructure than the model presented here because of the
higher density of independent tidal streams.

The predictions of WDM and broken scale-invariance models are less
clear, since they can reduce dwarf satellite numbers both by 
suppressing their formation in the first place (White \& Croft 2000)
and by making them less concentrated and therefore easier
to disrupt (Col{\'\i}n et al. 2000).
If the second effect dominates, the predictions might be closer
to those of SIDM; if the first effect dominates, they might be
closer to those of the reionization model.
At  present,  we  lack
quantitative predictions on  this matter from SIDM and  WDM, and we lack
detailed  observational  constraints, so we  cannot
draw conclusions about which model fares  best. 
However, it appears that studies of the mass, radial profile, and
substructure of the stellar halo might provide useful
constraints on these models in the near future.

It is interesting that all of the substructures detected in the nearly
all-sky  Carbon  star survey and the   SDSS  commissioning data
could be produced by a  single tidal stream  of the Sagittarius  dwarf
galaxy (Ibata et al. 2000b). In particular, the observed excess in the
RR Lyrae number density  profile at $r\approx  45$ kpc is caused  by a
clump of stars near the apocenter of the Sagittarius orbit. If most of the
RR Lyrae in the surveyed strip belong  to the Sagittarius stream, then
no stars  are expected beyond the  apocenter of the  Sagittarius orbit,
and this would naturally  explain the  drop in the RR  Lyrae number density at
$r\gsim 60$ kpc detected in the  SDSS data.  However, the presence of only
a single stream in a large volume of sky would be both puzzling and intriguing,
since the CDM models  considered in this  paper predict at least $\sim 20$
(and typically  more) tidal streams  in the halo of a Milky Way-size
galaxy. The  SIDM  and WDM stellar halos   should  be built  from even
larger number of tidal streams.  Carbon stars are relatively young and
rare,  and therefore older tidal streams  may  not be revealed by their
distribution.  We predict, however, that many more substructures
not associated with the Sagittarius tidal stream should be detected in the
future as the SDSS covers a larger area  of the sky.  Absence of such
detections would spell serious trouble for CDM models, and possibly even
more so for the variants of CDM that we have discussed.

Several past theoretical studies of tidal stripping and disruption 
of  galactic satellites have shown that
tidal streams can   be used as  powerful probes  of both   the present
potential of the Milky Way (Johnston et al.  1999; Ibata et al. 2000a)
and its accretion history  (e.g., Helmi \& White 1999;   Helmi et al.  
1999; Helmi \& de Zeeuw 2000).   
For example, Johnston   et al.   (1999) show that  parameters
characterizing the mass  distribution in the    Milky  Way  halo can    be
determined with  an  accuracy of a   few percent using  a single tidal
stream if accurate phase-space information is available for as few as 100
stream  stars.  Helmi et  al. (1999) show  how phase-space information
can be used to recover fossil remnants  of the satellites accreted and
disrupted early  in the Milky Way  evolution.  These techniques can be
used to construct a detailed formation history of the Galaxy.

In this study, we have focused on the spatial distribution of
stars in the Milky Way halo.  The radial density behavior
of stars may not
prove  to be as  sensitive a probe  of  the galactic  potential as the
accurate phase-space mapping of   tidal  streams (Johnston et al. 1999),   but
it should be very useful in recovering details of the Milky Way accretion
history.  In  some  respects, the number  of  surviving tidal streams or
the degree  of clumpiness of the  stellar  distribution can provide better
constraints  on the  variants of the 
CDM scenario than the abundance of
galactic satellites. An advantage here is that the spatial distribution of
halo stars  should be possible  to  map in the   very near future, when
large  samples of halo   stars from the SDSS and other surveys
become available.  Accurate,  large-scale, phase-space mapping of  tidal
streams, on the other hand, will  become possible only after launch of
the   next generation  astrometric   space missions (i.e., at the  end  of the
decade). On the theoretical side, our model  and predictions can  be
improved upon by combining   mass accretion histories typical for  the
galaxy-size halos formed    in  CDM models with  more    sophisticated
numerical models of orbital evolution of satellites and their tidal
debris.

\acknowledgements 

We thank  Amina Helmi and  Kathryn Johnston for  useful discussion and
suggestions.  This   work  was supported in  part   by NASA LTSA grant
NAG5-3525 and NSF grant AST-9802568.  Support for A.V.K.  was provided
by NASA through Hubble Fellowship grant HF-01121.01-99A from the Space
Telescope Science  Institute, which is  operated by the Association of
Universities  for Research  in  Astronomy, Inc.,  under  NASA contract
NAS5-26555.

\newpage

\end{document}